\documentclass[
    ,final            % use final for the camera ready runs
    ,sort&compress    % add further options here if necessary
  ]
  {aipproc}

\layoutstyle{6x9}

\def\aap{A\hbox{\rm \&}A} 
  \def\aj{AJ} 
  \def\apj{ApJ} \def\apjl{ApJL}
  
\def\araa{ARA\hbox{\rm \&}A}

\def\mnras{MNRAS} \def\nat{Nat}

\begin{document}

\title{Spectral energy distributions of submm/radio bright gamma-ray burst host galaxies}


\author{Micha{\l} J. Micha{\l}owski}{
  address={Dark Cosmology Centre, Niels Bohr Institute, University of Copenhagen, Juliane Maries Vej 30, DK-2100 Copenhagen, Denmark}
}

\author{Jens Hjorth}{
  address={Dark Cosmology Centre, Niels Bohr Institute, University of Copenhagen, Juliane Maries Vej 30, DK-2100 Copenhagen, Denmark}
}

\begin{abstract}

We present  optical to radio spectral energy distribution
fitting of the host galaxies of four long gamma-ray bursts: 980703, 000210, 000418 and 010222,
which were detected at submillimetre and/or radio wavelengths.
We find that only very young starburst galaxy models are consistent with the data
having both blue optical colors and a pronounced submm emission. For each host we are
able to construct a model consistent  with the short- and long-wavelength parts of the spectra. We find galaxy ages ranging from 0.09 to 2.0 Gyrs and star formation rates ranging  from  138 to 380 $M_\odot$ yr$^{-1}$.

\end{abstract}

\maketitle

%%%%%%%%%%%%%%%%%%%%%%%%%%%%%%%%%%%%%%%%%%%%
%% MAINMATTER
%%%%%%%%%%%%%%%%%%%%%%%%%%%%%%%%%%%%%%%%%%%%

\section{Introduction}

Observations of gamma-ray burst (GRB) host galaxies have attracted a lot of interest since
the first interpretation of a constant extended source at a GRB position as the host galaxy
\citep{vanparadijs,sahu97}. A detailed description of the GRB environment can provide
important constraints on GRB physics \citep[see][for a review]{hjorth04}
and can be used for statistical studies of starburst galaxies.

The host galaxies of the long GRBs 000210, 000418 and 010222 are somewhat special because they
are the only hosts, among  $\sim$30 targeted by SCUBA \citep{tanvir}, showing submillimetre emission.
The hosts of GRBs 980703 and 000418 were also firmly detected at radio wavelengths
\citep{frailwaxman,bergerkulkarni,berger} whereas those of GRBs 000210 and 010222 were only weakly detected
\citep{berger}. The host of GRB 010222 was also detected at millimetre wavelengths \citep{frail}.
Both submillimetre and radio detections may indicate high star formation rates (SFRs),
of the order of several hundreds solar masses per year ($M_\odot$ yr$^{-1}$),
if the whole emission is powered by a starburst, namely by dust at submillimetre and 
by supernova remnants at radio wavelengths.
Moreover, a spectral energy distribution (SED) fitting applied only to optical/near-infrared data agreed reasonably
with starburst templates \citep{gorosabel1,gorosabel2,christensen04}.
This picture is consistent with the hypernova GRB model \citep{woosley,paczynski98}.

However, the starburst scenario of GRB hosts is more complicated because,
unlike red, dusty, sturbursting submillimetre galaxies \citep{smail04}, 
GRB hosts have been found to exhibit blue optical colors \citep{vreeswijk,gorosabel1,gorosabel2,galama}. 
The submillimetre/radio fluxes
were also underestimated by any SED template
fitted  to the optical data. 

Mid-infrared {\it Spitzer} observations of the hosts of GRBs 980703 and 010222 could not bridge the optical and submillimetre
data and provide an explanation of the unusual SED behaviour and a clue on dust properties because only the former was detected
in the bluest $4.5\,\mu$m {\it Spitzer} passband \citep{lefloch,castroceron06}.

In this contribution we attempt to find reasonable full SED fits for those four GRB host galaxies
and constrain their ages and SFRs. This approach seems to be a promising tool to constrain the dust properties
and the obscured star formation rates as shown, for example, by \citet{priddey06}.
We use a cosmological model characterized by $H_0=70$ km s$^{-1}$ Mpc$^{-1}$, 
$\Omega_\Lambda=0.7$ and $\Omega_M=0.3$.

\section{SED modelling}

In order to model the SEDs of GRB hosts we used the GRASIL\footnote{\url{http://web.pd.astro.it/granato/grasil/grasil.html}}
software described by \citet{silva98}.
It is a numerical code that calculates the spectrum of
a galaxy by means of a radiative transfer method, applied to photons produced by a stellar population,
and reprocessed by dust. The importance of this model is the fact that it is self-consistent
in that it fulfills the principle of energy conservation between the energy absorbed by dust in the UV/optical wavelengths
and the energy re-emitted in the infrared.

\begin{figure}[p]
\includegraphics[width=\textwidth,%viewport=35 245 570 750, clip
]{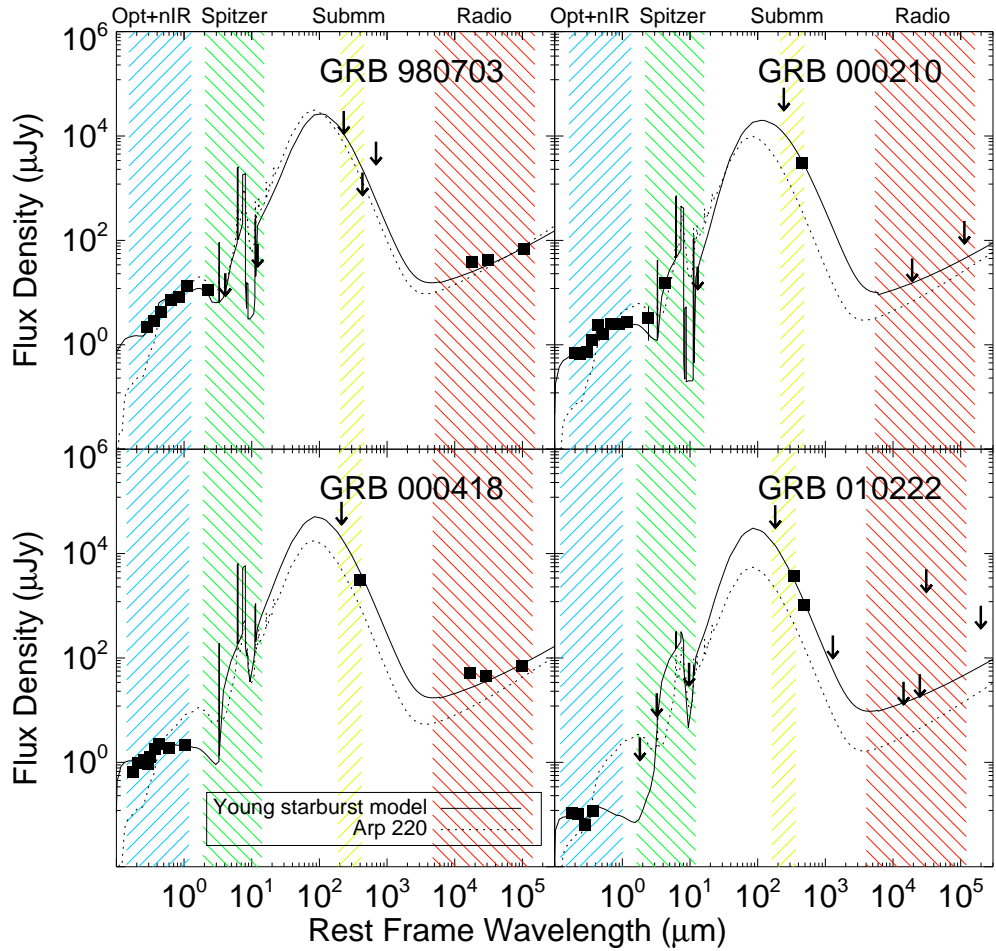}
\caption{SEDs of GRB hosts. Solid lines: the young starburst galaxy models calculated using GRASIL and consistent with the data. Dotted lines: Arp 220 model \citep[from][]{silva98}.
Squares: detections with errors, in most cases, smaller than the size of the symbols. Arrows: $3\sigma$ upper limits (values marked at the base).
Hashed columns mark the wavelengths corresponding
to optical/near infrared filters, mid-infrared Spitzer filters, SCUBA submillimetre bands and the radio domain. Data are from:
optical/near-infrared: \citet{sokolov01, vreeswijk,gorosabel1,gorosabel2,galama}; mid-infrared: \citet{castroceron06,lefloch};
submillimetre: our  reduction of archival SCUBA data, \citet{tanvir}; radio: \citet{bergerkulkarni,berger,frail,sagarstalin01}
 }
\label{fig:myseds}
\end{figure}

Figure \ref{fig:myseds} shows that  submm and radio emissions of the GRB hosts discussed here are underestimated even for a template corresponding to the nearby
ultraluminous infrared galaxy (ULIRG) Arp 220 \citep{silva98}, which is also too red in the optical compared to GRB hosts. This indicates the need
for new empirical SED models, which are presented here.

We performed  SED modelling investigating a wide range of GRASIL parameters
(see \citet{michalowskimaster} for a description)
We scaled the SED templates to the datapoints
and chose
those which resulted in acceptably small $\chi^2$.
Then we calculated the total SFRs
using the infrared luminosity, integrating between 8 and 1000 $\mu$m (rest frame), and applying
 \citet{kennicutt}:
\mbox{SFR$(M_\odot\mbox{ yr}^{-1})=4.5\cdot10^{-44}L_{8-1000}$(erg s$^{-1}$)}.

%in order to be able to produce multiple footnotes using \footnotemark
\renewcommand{\thefootnote}{\fnsymbol{footnote}}

\begin{table}
\begin{tabular}{l r@{.}l r@{.}l r@{ $\pm$ }l l c r@{ $\pm$ }l  l}
\hline
\tablehead{1}{l}{b}{GRB\\ host}
&\tablehead{2}{c}{b}{$\bf z$}
&\tablehead{2}{c}{b}{Age\tablenote{Defined as the time since the beginning of the galaxy evolution.}\\ (Gyr)}
&\tablehead{6}{c}{b}{SFR\\ ($\bf M_\odot$ yr$\bf^{-1}$)}
&\tablehead{1}{c}{b}{References}\\
\cline{6-11}
\multicolumn{5}{c}{}
& \tablehead{2}{c}{b}{This work\tablenote{Errors are statistical at $1\sigma$ level calculated assuming that the template fits to the data.}}
\addtocounter{mpfootnote}{1}     % to avoid ** which doesn't fit
& \tablehead{1}{c}{b}{Infrared}
& \tablehead{1}{c}{b}{Submm}
& \tablehead{2}{c}{b}{Radio}
&\\
\hline
980703& 0&97&  2&0 & 179&29 & <24\tablenote{Based only on $24\,\mu$m \citep{lefloch}}, <226\tablenote{All photometric datapoints taken into account \citep{castroceron06}}
	 & <380 & 180&25 & \citep{djorgovski98}, \citep{lefloch}, \citep{castroceron06}, \citep{berger}\\
000210& 0&85 & 0&3 & 138&17 & & 560 $\pm$ 165 & 90&45 & \citep{piro}, \citep{berger}\\
000418& 1&12&  0&14& 380&46 & & 690 $\pm$ 195 & 330&75 & \citep{bloom03b}, \citep{berger} \\
010222& 1&48& 0&09 & 366&84 & <130\footnotemark[4] & 610 $\pm$ 100 & 300&115 &  \citep{jha01}, \citep{lefloch}, \citep{berger}\\
\hline
Arp 220\tablenote{The Arp~220 parameters given for the comparison.}
& 0&02& 13&0 & \multicolumn{1}{l}{580} &&&& \multicolumn{2}{c}{} & \citep{arp220}, \citep{silva98}\\
\hline
\end{tabular}
\caption{Ages and total SFRs of GRB host galaxies. All four are young starbursts.}
\label{tab:grasilparam}
\end{table}

\clearpage

\section{Results and Discussion}

For each host we were able to construct an SED that fitted the data reasonably well. 
The corresponding ages of the galaxies and the SFRs 
are presented in Table~\ref{tab:grasilparam} and fits are shown in Figure~\ref{fig:myseds}.

All the galaxies discussed are young (ages less than 2 Gyrs).
Here, the age of a galaxy was defined as the  time since the beginning of its evolution, when the stellar population
starts to build up.
The bigger the difference between optical and submillimetre fluxes, the lower the needed age was.
This is because a younger galaxy has more stars still embedded in molecular clouds, so optical light is weaker and dust emission stronger.

The construction of the SEDs assuming low ages for the galaxies was therefore the way to take into account both
significant submillimetre/radio emission and the blue optical colors of GRB hosts.
In young galaxies  there are, on one hand,  lots of
young, hot, blue stars, because they have not finished their lives yet.
Hence the total optical spectrum of the galaxy
is blue. On the other hand, the majority of the stars still resides in dense molecular clouds,
so a significant part of the energy is absorbed and re-emitted. 
This increases the dust emission. 
GRBs reside in molecular clouds and it was  found that gas column densities
derived from $X$-ray afterglows in a sample of 8 GRBs (including GRB 980703, discussed here) were in the range
corresponding to the column densities of giant molecular clouds in the Milky Way \citep{galamawijers01}.
A similar conclusion for high-redshift GRBs was recently drawn by \citet{jakobsson06b} by means of modelling Ly$\alpha$ absorption features.

Based on SEDs fitted to optical data, \citet{gorosabel1,gorosabel2} and \citet{christensen04}
derived the ages of the starbursts in the hosts of GRBs 980703, 000210, 000418.
Our estimates are at most a factor of two larger. Given that we calculated the time
from the beginning of the galaxy evolution, not the beginning of the starburst,
our values agree with those derived by these
authors. \citet{sokolov01} derived ages of both old stellar populations and starbursts.
Our estimation for GRB 980703 agrees
with the age of the old component, which is conceptually closer to our definition
of the galaxy age.

Table \ref{tab:grasilparam} also presents the SFRs derived from the SED fits.
They are all  high, of the order of several hundreds solar masses per year.
Statistical $1\sigma$ errors, 
calculated assuming that the model represents the data, are also given in Table \ref{tab:grasilparam}.
On top of that, one should add the 30\% uncertainty of the relation between infrared luminosities and SFRs \citep{kennicutt}.

In Table \ref{tab:grasilparam} our SFR estimations are compared with those derived using three other
SFR indicators: mid-infrared, submillimetre and radio emission.
Our results are consistent with radio-derived SFRs \citep{berger}. This is because the calibration of SFR to
radio flux requires choosing only two parameters  \citep[a normalization factor and a spectral index, see][]{yun},
which are relatively well constrained. Our values also agree with the upper limits derived by \citet{castroceron06}
using the template of Arp 220. This approach appears to be more self-consistent and reliable than analysing
only $24\,\mu$m datapoints, which is strongly dependent on the unknown shape of the infrared part of SED.
For example, this method applied by \citet{lefloch} gave upper limits inconsistent with our results.
Finally, \citet{berger} obtained higher SFRs based on submm alone.

The mid-infrared ($5\,\mu\mbox{m}<\lambda<40\,\mu\mbox{m}$) shape of the SEDs
was not constrained due to a lack of targeted observations
or detections. Hence, all SEDs shown on Figure \ref{fig:myseds} have different
mid-infrared behaviour depending on type, mixture and sizes of dust grains used.
Any data in this region would be useful to learn more about the dust properties
in these galaxies.

This work provides strong evidence that some GRB hosts are young starbursts.
We have proposed an explanation why they have both
blue optical colors and significant submillimetre emission 
unlike red, dusty submillimetre galaxies \citep{smail04}. 
The SED approach is self-consistent and makes use of all the data available. 
The research presented here will be continued to build up a larger galaxy sample analysed in this way.
Other properties of the host galaxies, the robustness of the technique and 
a more accurate treatment of the errors are discussed in \citet{michalowski07,michalowski07b}.

%%%%%%%%%%%%%%%%%%%%%%%%%%%%%%%%%%%%%%%%%%%%%%%%
%% BACKMATTER
%%%%%%%%%%%%%%%%%%%%%%%%%%%%%%%%%%%%%%%%%%%%%%%%

\begin{theacknowledgments}
We thank Joanna Baradziej, Jos\'{e} Mar\'{i}a Castro Cer\'{o}n,
Darach Watson, Frank Bertoldi and Mike Garrett for useful discussion and comments,
Brad Cavanagh,  Frossie Economou, Tim Jenness and Carsten Skovmand for help with the \verb+ORAC-DR+
installation. MM thanks the organizers of the conference for a very inspiring meeting.
The Dark Cosmology Centre is funded by the Danish National Research Foundation.
\end{theacknowledgments}

%%%%%%%%%%%%%%%%%%%%%%%%%%%%%%%%%%%%%%%%%%%%%%%%
%% The bibliography can be prepared using the BibTeX program or
%% manually.
%%
%% The code below assumes that BibTeX is used.  If the bibliography is
%% produced without BibTeX comment out the following lines and see the
%% aipguide.pdf for further information.
%%
%% For your convenience a manually coded example is appended
%% after the \end{document}
%%%%%%%%%%%%%%%%%%%%%%%%%%%%%%%%%%%%%%%%%%%%%%%%

%%%%%%%%%%%%%%%%%%%%%%%%%%%%%%%%%%%%%%%%%%%%%%%%
%% You may have to change the BibTeX style below, depending on your
%% setup or preferences.
%%
%%
%% For The AIP proceedings layouts use either
%%%%%%%%%%%%%%%%%%%%%%%%%%%%%%%%%%%%%%%%%%%%

%\bibliographystyle{aipproc}   % if natbib is available
%\bibliographystyle{aipprocl} % if natbib is missing

%%%%%%%%%%%%%%%%%%%%%%%%%%%%%%%%%%%%%%%%%%%
%% You probably want to use your own bibtex database here
%%%%%%%%%%%%%%%%%%%%%%%%%%%%%%%%%%%%%%%%%%%
%\bibliography{michalowski}

%%%%%%%%%%%%%%%%%%%%%%%%%%%%%%%%%%%%%%%%%%%
%% Just a reminder that you may have to run bibtex
%% All of it up to \end{document} can be removed
%% if you don't like the warning.
%%%%%%%%%%%%%%%%%%%%%%%%%%%%%%%%%%%%%%%%%%%
\IfFileExists{\jobname.bbl}{}
 {\typeout{}
  \typeout{******************************************}
  \typeout{** Please run "bibtex \jobname" to optain}
  \typeout{** the bibliography and then re-run LaTeX}
  \typeout{** twice to fix the references!}
  \typeout{******************************************}
  \typeout{}
 }

\end{document}